\documentclass[aps,prd,twocolumn,nofootinbib,showpacs,floatfix,preprintnumbers,groupedaddress]{revtex4}
\usepackage{graphicx}
\usepackage{epsfig}

\newcommand{\sla}[1]{/\!\!\!#1}
\newcommand{\Sla}[1]{/\!\!\!\!#1}
\def\lsim{\raise0.3ex\hbox{$\;<$\kern-0.75em\raise-1.1ex\hbox{$\sim\;$}}}
\def\gsim{\raise0.3ex\hbox{$\;>$\kern-0.75em\raise-1.1ex\hbox{$\sim\;$}}}

\def\hbar{\hspace{0pt}\raisebox{1pt}{$-$} \hspace{-7pt} h}

\newcommand {\ignore}[1]{}

\begin{document}
\preprint{YITP-SB-08-44}

\title{Deciphering the spin of new resonances in Higgsless models}

\author{Alexandre Alves}
\email{aalves@fma.if.usp.br}
\affiliation{Instituto de F\'{\i}sica,
             Universidade de S\~ao Paulo, S\~ao Paulo -- SP, Brazil.}

\author{O.\ J.\ P.\ \'Eboli}
\email{eboli@fma.if.usp.br}
\affiliation{Instituto de F\'{\i}sica,
             Universidade de S\~ao Paulo, S\~ao Paulo -- SP, Brazil.}

\author{M.\ C.\ Gonzalez--Garcia} \email{concha@insti.physics.sunysb.edu}
\affiliation{%
  Instituci\'o Catalana de Recerca i Estudis Avan\c{c}ats (ICREA),
  Departament d'Estructura i Constituents de la Mat\`eria, Universitat
  de Barcelona, 647 Diagonal, E-08028 Barcelona, Spain}
\affiliation{%
  C.N.~Yang Institute for Theoretical Physics, SUNY at Stony Brook,
  Stony Brook, NY 11794-3840, USA}

\author{J.\ K.\ Mizukoshi} \email{mizuka@ufabc.edu.br}
\affiliation{Centro de Ci\^encias Naturais e Humanas, Universidade Federal do 
ABC, Santo Andr\'e -- SP, Brazil.}


\begin{abstract}
\vspace*{1cm}

We study the potential of the CERN Large Hadron Collider (LHC) to
probe the spin of new massive vector boson resonances predicted by
Higgsless models. We consider its production via weak boson fusion
which relies only on the coupling between the new resonances and
  the weak gauge bosons.  We show that the LHC will be able to
unravel the spin of the particles associated with the partial
restoration of unitarity in vector boson scattering for integrated
luminosities of 150--560 fb$^{-1}$, depending on the new state
mass and on the method used in the analyses.

\end{abstract}

\maketitle



\section{Introduction}

Despite the success of the Standard Model (SM) of particle physics in
describing electroweak physics below $\sim$ 100 GeV in terms of a
non-abelian gauge theory with spontaneously broken $SU(2)_L\times
U(1)_Y$ gauge group, the gauge symmetry does not predict the precise
mechanism of the electroweak symmetry breaking (EWSB). Indeed, up to
this moment, there is no direct experimental signal of the mechanism of
EWSB, being its search one of the main goals of the LHC.

The EWSB mechanism plays an important role in the high energy electroweak
gauge boson scattering which violates partial wave unitarity or becomes
strongly interacting at energies of the order of $E\sim 2$ TeV, if there is no
new state to cut off its growth ~\cite{Lee:1977yc,Lee:1977eg}.  In the context
of the SM, as well as in its supersymmetric realization, electroweak symmetry
is broken by the vacuum expectation value of some weakly coupled neutral
scalar state(s), the Higgs boson(s), which will contribute to electroweak gauge
boson scattering, preventing the unitarity violation of the process.

Alternatively, {\sl Higgsless} extensions of the
SM~\cite{Csaki:2003dt,Csaki:2003zu,Nomura:2003du} have been proposed
in which the electroweak symmetry is broken without involving a
fundamental Higgs field.  Generically on these models, the electroweak
symmetry is broken by boundary conditions in
a 
higher dimensional space. The originally proposed Higgsless models
gave large contributions to precision electroweak observables, in
particular to the $S$ parameter~\cite{Peskin:1991sw}
($\epsilon_3$~\cite{Altarelli:1993sz})~\cite{Barbieri:2003pr,Burdman:2003ya,Davoudiasl:2003me,Cacciapaglia:2004jz,Davoudiasl:2004pw,Barbieri:2004qk}.
Such problems could be overcome, for example, by appropriate
modifications of the fermion sector. In this way, a variety of
Higgsless models have been
constructed~\cite{Foadi:2003xa,Hirn:2004ze,Cacciapaglia:2004rb,
  Casalbuoni:2004id,
  Chivukula:2004pk,Foadi:2004ps,Georgi:2004iy,Casalbuoni:2005rs,SekharChivukula:2005cc,
  Chivukula:2005ji,Cacciapaglia:2006gp} which ensure agreement with
electroweak precision data.

From the point of view of unitarity, all Higgsless models share the
common feature that new weakly interacting spin-1 gauge bosons
particles with the same quantum numbers as the SM gauge bosons appear
and they are responsible for the partial restoration of unitarity in
vector boson scattering and for rendering a theory weakly
coupled to energies well above 2
TeV~\cite{SekharChivukula:2001hz,Chivukula:2003kq,SekharChivukula:2008mj}.
This property allows for an almost model independent search for the
lightest charged resonance $V_1^\pm$ at the LHC through $pp \to
V_1^\pm W^\mp$ or via weak boson fusion $pp \to V_1^\pm q
q$~\cite{Birkedal:2004au,He:2007ge}, as long as $V_1^\pm$ remains
  a narrow resonance. The LHC experiments will be able to unravel the
existence of the charged state via these processes with modest
integrated luminosities of 10--60 fb$^{-1}$. On the contrary, the
corresponding search for the neutral vector resonance in gauge boson
fusion is expected to be very difficult, since a generic feature of
this class of models is the absence of coupling between the neutral
resonance and $ZZ$ pairs.  Reconstructing the heavy neutral vector
resonance decaying into $W^+W^-$ requires at least one hadronic $W$
decay, posing the challenge to dig it out from the large SM
backgrounds.

Once a clear signal of the charged resonance is observed in the above
channels, it is mandatory to study its spin to confirm that the new
state is indeed a vector particle. In this work, our goal is to probe
the $V_1^\pm$ spin via the study of weak boson fusion production of
$V_1^\pm$ with its subsequent decay into leptons, {\em i.e.}
\begin{equation}
   pp \to V_1^\pm jj \to \ell^\pm \nu \ell^{\prime+} \ell^{\prime-} jj 
\label{eq:procsignal}
\end{equation}
with $\ell$ and $\ell^\prime = e$ or $\mu$, considering final states where
  the $W$'s and $Z$'s decay into different and same flavor charged leptons.
To determine the spin of the state decaying into $W^\pm Z$ we contrast
the final state distributions arising from the production and decay of
the vector charged state with the ones stemming from the decay of a
scalar state; {\em i.e.}  we work in the framework commonly used to
analyze the spin of supersymmetric
particles~\cite{susyspin1,susyspin2}. Here we show that it is possible
to determine the spin of a new heavy resonance decaying into $W^\pm Z$
at the LHC with 99\% CL for luminosities of $\sim$ 150--560
fb$^{-1}$, depending on the particle mass and the method used in the
analysis.

\section{Model and Calculation Setup}
\label{model}

The restoration of partial wave unitarity in Higgsless models is due
to new Kaluza--Klein resonances $V_{(i)}^\pm$ and $V_{(i)}^0$. The
couplings $V_{(i)}^+ W^- Z$ ($V_{(i)}^0 W^- W^+$) have the same
Lorentz structure of the SM $W^+ W^- Z$ vertex with a coupling constant
$g^{(i)}_{VWZ}$ ($g^{(i)}_{VWW}$). In order to cancel the dangerous
terms in the scattering $WZ \to WZ$ that depend on $E^2$ and $E^4$,
where $E$ is the energy of the incoming $W$ and $Z$ in the
center--of-mass system, the new vector state coupling constants must
satisfy the following constraints:
\begin{eqnarray}
&&  g_{\text{WWZZ}} = 
            g_{\text{ZWW}}^2  + \sum_i \left(g^{(i)}_{\text{VWZ}} \right)^2\;,
\label{const1}
\\
 && 2\left ( g_{\text{WWZZ}} - g_{\text{ZWW}}^2 \right) \left(M_W^2+M_Z^2\right) 
+ g^2_{\text{ZWW}} \frac{M_Z^4}{M_W^2} \nonumber 
\\&&=
 \sum_i   \left(g^{(i)}_{\text{VWZ}} \right)^2 
\left[ 
      3 \left( M_i^\pm\right)^2 - \frac{ (M_Z^2-M_W^2)^2}{\left(M_i^\pm\right)^2}
\right]\;.
\label{const2}
\end{eqnarray}

Eqs.\ (\ref{const1}) and (\ref{const2}) constrain the couplings of the
lightest charged Kaluza-Klein state to $WZ$ pairs,
\begin{equation}
    g^{(1)}_{\text{VWZ}} \lesssim \frac{g_{\text{ZWW}} M^2_Z}{\sqrt{3} M_1^\pm
    M_W} \; .
\end{equation}
In our analysis we assume that this bound is saturated 
\cite{Cheung:2007in}, which leads to
the largest allowed value for $g^{(1)}_{\text{VWZ}}$, and we evaluate
the quartic coupling $g_{\text{WWZZ}}$ using Eq.~(\ref{const1}).
  Moreover, we assume that the $V_1^\pm$ couplings to fermions are
  small and that the $V_1^\pm$'s mainly decay into $WZ$ pairs.  This
  hypothesis is, in fact, realized in some higgsless
  models~\cite{Cacciapaglia:2006gp}.


Our study of the $V_1^\pm$ spin was carried out by comparing the kinematic
distributions of its decay products with the predictions for the production of
a spin-0 resonance. Since the signal for the new charged state is
characterized by peak in the $WZ$ invariant mass distribution, we use as
template the kinematic distributions in a model which is the SM without a
Higgs plus a scalar charged state, $H^\pm$, with an interaction $H^\pm Z^\mu
W^\mp_\mu$.  The coupling of the $H^\pm Z^\mu W^\mp_\mu$ vertex is chosen such
that the $H^\pm$ production cross section is equal to the one for $V_1^\pm$
after all cuts. We also set the $H^\pm$ width equal to the $V_1^\pm$ one.

We performed a parton level study using the full tree level amplitude for the
final state processes in order to keep track of spin correlations. The matrix
elements were generated using the package MADGRAPH~\cite{madevent}, where we
included the higgsless (and template) model particles and interactions. We
employed the CTEQ6L parton distribution functions~\cite{Pumplin:2002vw} with
the factorization scale $\mu_F = \sqrt{(p^2_{T~j1} + p^2_{T~j2} )/2 } $, where
$p_{T~ji}$ are the transverse momenta of the tagging jets. For the QCD
backgrounds we chose the renormalization scale $\mu_R=\mu_F$. In order to have
a crude simulation of the detector performance we smeared energies, but not
directions, of all final state partons with a Gaussian error. For the
jets, we assumed a resolution  $\Delta E/E = 0.5/\sqrt{E} \oplus 0.03$, 
if $|\eta_j| \leq3$, and $\Delta E/E = 1/\sqrt{E} \oplus 0.07$, 
if $|\eta_j| >3$ ($E$ in GeV), while for charged leptons we used a
resolution  $\Delta E/E = 0.1/\sqrt{E} \oplus 0.01$.  Furthermore, we 
considered the jet tagging efficiency to be $0.75 \times 0.75=0.56$, while
the lepton detection efficiency is taken to be $0.9^3 =0.73$.

\section{Results}
\label{results}

We analyzed the process
\[
pp \to  \ell^\pm  \ell^{\prime +} \ell^{\prime -} jj+ \Sla{E_T} 
\;\; \hbox{ with } \ell,~\ell^\prime = e,~\mu
\] 
which contains the contribution of the vector boson fusion
production of new charged resonances decaying into leptons; see
Eq.~(\ref{eq:procsignal}).
This process possesses a significant irreducible background
originating from electroweak and QCD $WZjj$ production. Moreover, the
production of $t \bar{t}$ pair in association with a jet exhibits a
large cross section after we demand the presence of two tagging
jets~\cite{Eboli:2006wa} and can lead to trilepton events when both
$t$'s decay semi-leptonically and the decay of one of the $b$'s leads
to an isolated lepton\footnote{We considered a lepton to be isolated
  if the hadronic energy deposited in a cone of size $\Delta R < 0.4$
  is smaller than 10 GeV}.

Initially we imposed the following jet acceptance cuts designed
  to enhance events produced by vector boson fusion,
\begin{equation}
 \begin{array}{c}
 p_T^j > 20 \hbox{ GeV} \;\;\;\;\;\;\; , \;\;\;\; | \eta_j | < 4.9 \;\; ,
 \\
 | \eta_{j1} - \eta_{j2} | > 3.8 \;\;\; , \;\;\;
 \eta_{j1} \cdot \eta_{j2} < 0 \;\; .
 \end{array}
\label{cuts1}
\end{equation}

We also applied lepton acceptance and isolation cuts
\begin{equation}
\begin{array}{cc}
|\eta_\ell|\leq 2.5 \;\;\; &, \;\;\;
p_{T}^\ell \geq 10 \; {\rm GeV}
\\
\Delta R_{\ell j}\geq 0.4 \;\;\; &, \;\;\;
\Delta R_{\ell \ell}\geq 0.4 \;\; .
\nonumber
\end{array}
\label{cuts2}
\end{equation}

\onecolumngrid

\begin{table}[t]
 \begin{tabular}{|l||c|c|c|c|}
\hline \hline
         
&  cuts (\ref{cuts1})--(\ref{cuts2})
&  cuts (\ref{cuts1})--(\ref{cuts2a})
&  cuts (\ref{cuts1})--(\ref{cuts2b})
&  cuts (\ref{cuts1})--(\ref{cuts3})
\\
\hline
EW $WZjj$ & 4.68 & 2.68 & 2.40 &  0.265/0.166
\\
\hline
$t \bar{t}j$ & 22.4 & 6.54  & 1.85 & 0.024/0.0003 
\\
\hline
$M_{V^\pm_1}= 500$ GeV & 1.02  & 0.84 & 0.78  &  0.705 
\\
\hline
$M_{V^\pm_1}= 700$ GeV & 0.36  & 0.32 & 0.30 &  0.25 
\\
\hline \hline
 \end{tabular}
 \caption{SM background and signal cross sections after different set of cuts
   in fb. In the last column the top/bottom results is obtained applying the 
   top/bottom cut in Eq.~(\ref{cuts3}).}
 \label{tab:disc}
\end{table}
\twocolumngrid
 
As we can see from Table~\ref{tab:disc} the SM background is
  still quite large after these cuts 
  with the $t \bar{t} j$ production being the dominant
  contribution. In order to reduce this background we explore two
  features of the signal and backgrounds. First of all, in the $t
  \bar{t}j$ production the lepton coming from the $b$ semi-leptonic
  decay is quite soft, therefore, it can be reduced by imposing an
  additional lepton transverse momentum cut:
\begin{equation}
   p_T^\ell > 25 \hbox{ GeV.}
 \label{cuts2a}
\end{equation}
Moreover, two of the leptons in the signal come from a $Z$ decay,
consequently we also required that the events present a pair of same
flavor opposite charge leptons (SFOC) with an invariant mass in a
window around the $Z$ mass. Thus we further demanded
\begin{equation}
   \mid M_{\ell\ell}^{\rm SFOC}   - M_Z \mid < 10 \hbox{ GeV.}
\label{cuts2b}
\end{equation}

\begin{figure}[!t]
\centerline{\psfig{figure=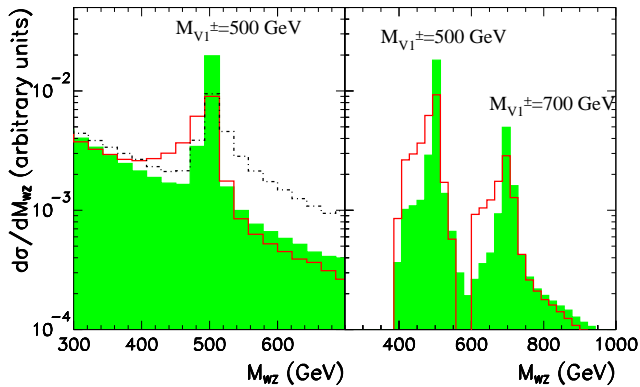,width=0.5\textwidth}}
\caption{{\bf Left :} $WZ$ invariant mass distribution reconstructed using the
  largest (dash-dotted line) and smallest (full line) estimate [see
  Eq.\ (\ref{eq:plnu})], as well as the true distribution (shadowed region) 
  for two values of $M_{V_1^\pm}$.  {\bf Right:} The true (shadow) and minimum 
  (line) reconstructed $WZ$ invariant mass distributions after the cuts in
  Eq.\ (\ref{cuts3}).}

\label{fig:mwz}
\end{figure}

The presence of just one neutrino in the signal final state,
Eq.~(\ref{eq:procsignal}), allows for full reconstruction of the neutrino
momentum -- up to a twofold ambiguity on its longitudinal component -- by
imposing the transverse momentum conservation and requiring that the invariant
mass of the neutrino--$\ell^\pm$ pair, where $\ell$ is the charged lepton
not identified as coming from the $Z$ decay,
%
%
is compatible with the $W$ mass:
\begin{eqnarray}
p_L^\nu&=&\frac{1}{2 {p^l_T}^2}
\bigg\{\big[M_W^2+2(\vec {p^l_T} \cdot \vec{\sla{p_T}})\big] p_L^l \nonumber \\
&&\pm 
\sqrt{\big[M_W^2+2(\vec{p_T^l} \cdot \vec{\sla{p_T}})\big]^2 |\vec p^l|^2 -
4 (p_T^l E^l\Sla{E_T})^2}\bigg\}
\label{eq:plnu}
\end{eqnarray}
Consequently, there are two distinct estimates for the $WZ$ invariant mass
which we label $M^{\rm rec,max}_{WZ}$ and $M^{\rm rec,min}_{WZ}$ the maximum
and minimum reconstructed values, respectively.
We show in left panel of Fig.\ \ref{fig:mwz} the $M^{\rm rec,max}_{WZ}$ and
$M^{\rm rec,min}_{WZ}$, as well as the true $M_{WZ}$ invariant mass
distributions for $M_{V_1^\pm}= 500$ GeV.  As seen in the figure, both
reconstructed distributions present a clear peak associated to the presence of
a new charged resonance. Moreover, the maximum (minimum) reconstructed $WZ$
invariant mass is a reasonable estimator of the true distribution for $WZ$
invariant masses smaller (larger) than the position of the resonance.

In order to isolate the contribution of the new charged states, we
imposed a cut on  $M^{\rm rec,min}_{WZ}$ 
\begin{equation}
  \begin{array}{ll}
 400 \hbox{ GeV} \le M^{\text{rec, min}}_{WZ}  \le 550 \hbox{ GeV}, 
& \hbox{ for   }   M_{V_1^\pm} = 500 \hbox{ GeV}
\\
\\
  600 \hbox{ GeV} \le M^{\text{rec, min}}_{WZ},  
& \hbox{ for   } M_{V_1^\pm} = 700 \hbox{ GeV}
  \end{array}
\label{cuts3}
\end{equation}
The effect of these cuts on the $WZ$ invariant mass spectrum can be seen in
the right panel of Fig.\ \ref{fig:mwz} for $M_{V_1^\pm}= 500$ GeV and 700 GeV.
As seen in the figure after these cuts, a good fraction of the peak signal
events are retained.

The predicted cross sections for the signal and SM backgrounds after
cuts (\ref{cuts1})--(\ref{cuts3}) are listed in Table~\ref{tab:disc}.
From these numbers we conclude that the above cuts lead to a good
  signal to background ratio of $\simeq 2.4$ (1.5) for $M_{V^\pm_1} =
  500$ (700) GeV.  Thus, a clear observation ($5 \sigma$) of the new charged
  resonances $V_1^\pm$ with a 500 (700) GeV mass in the leptonic
  channel requires a modest integrated luminosity of 15 (66)
  fb$^{-1}$, which can be achieved in the low luminosity run of the
  LHC or in the early stages of the high luminosity run.

Similar sensitivity could be obtained by cutting, instead, on $M^{\rm
  rec,max}_{WZ}$, though in general the cuts have to be chosen
  tighter and dependent on the $M_{V_1^\pm}$ mass.  This is so
because the SM background is a decreasing function of the $WZ$ mass,
therefore when cutting on the maximum reconstructed $WZ$ mass, the
number of miss--reconstructed background events in the signal region
tends to be larger.

After the new state coupled to $WZ$ is discovered, it is important to probe
its spin. The best way to accomplish that is to study angular correlations of
the final state particles. In principle, useful information on the spin could
be also extracted from the production cross section, however, at the LHC one
measures the production cross section times the decay branching ratio,
requiring additional information to disentangle these quantities.  Here we
employ two methods to unravel the spin of the new charged state based
exclusively on the kinematic distribution of the final state particles.  In
the first method, we contrast the kinematic distributions of the charged
leptons produced in the decay of vector and scalar charged states, much in the
spirit of the analysis carried out to study the spin of supersymmetric
particles at the LHC~\cite{susyspin1,susyspin2}.  A virtue of this method is
that it does not rely on the reconstruction of the neutrino momentum (besides
the invariant mass cut). In our second analysis, we used the reconstructed
neutrino momentum to obtain the polar angle of the produced $Z$'s in the $WZ$
center--of--mass system.

\begin{figure*}[!t]
       \centerline{\psfig{figure=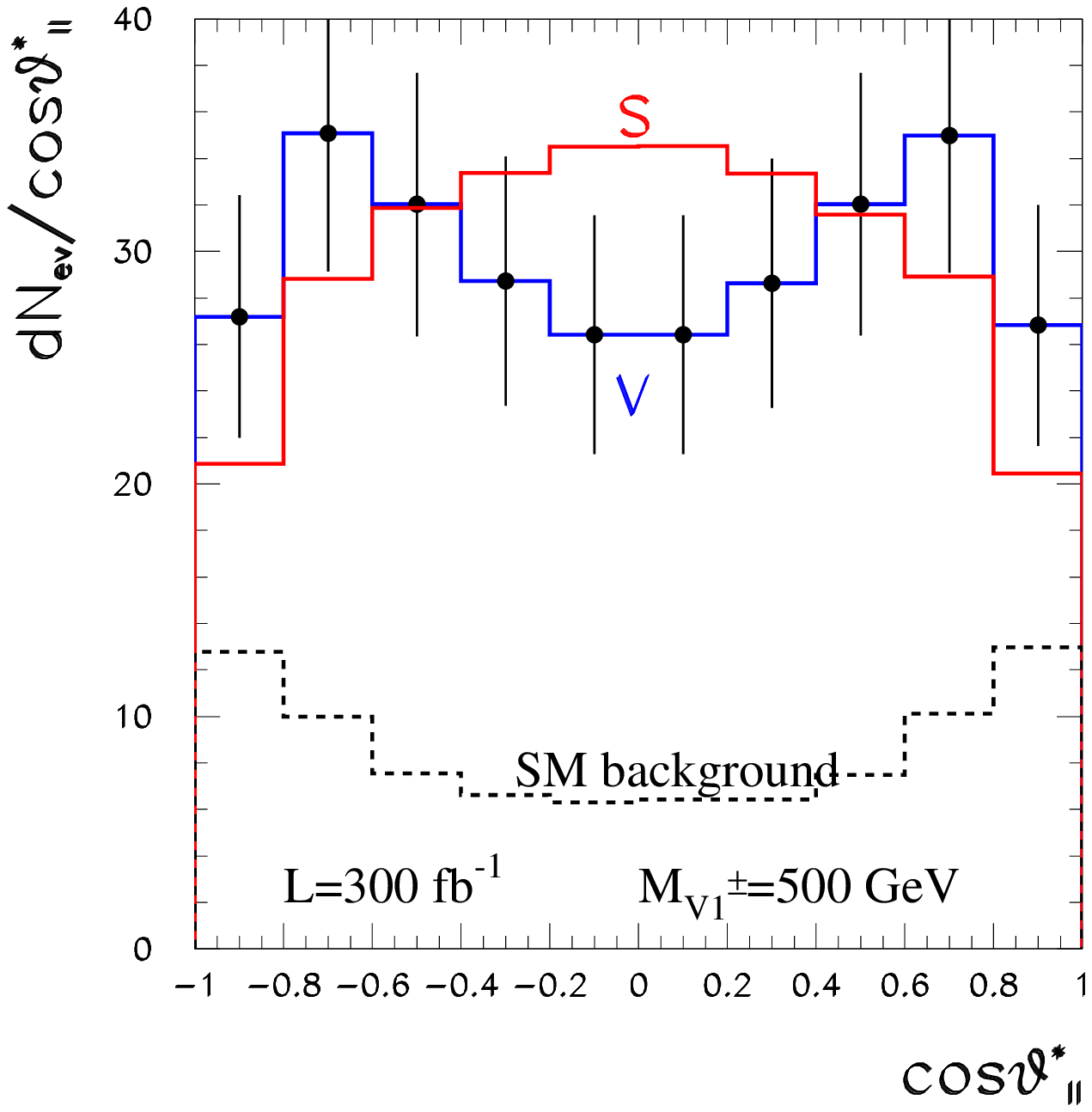,width=0.55\textwidth,
height=0.3\textheight}\hspace*{-1.5cm}
                   \psfig{figure=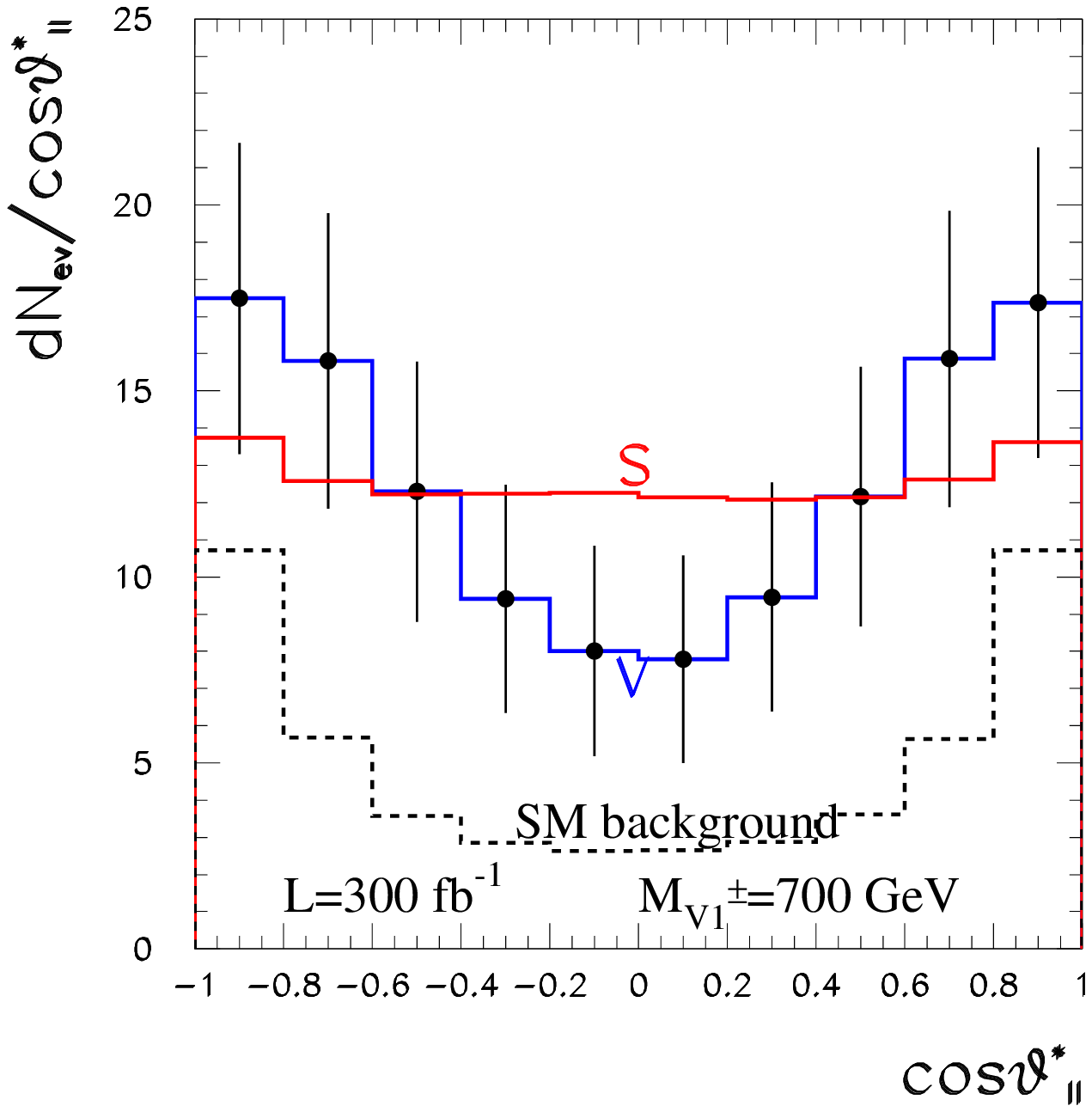,width=0.55\textwidth,
height=0.3\textheight}}
\caption{$\cos\theta^*_{\ell\ell}$ distribution for the SM background
  (dashed line), the production of a vector charged resonance (solid
  line with error bars), and the production of charged scalars (solid
  line). In the left (right) panel the mass of the new resonance is
  500 (700) GeV and we considered an integrated luminosity of 300 fb$^{-1}$.}
\label{fig:barr}
\end{figure*}

In order to contrast the spin-0 and spin-1 resonances, we focused on the
leptons whose momenta can be well determined. In previous
studies~\cite{susyspin1}, it has been shown that a convenient variable for such
analysis is
\begin{equation}
   \cos\theta_{\ell\ell}^*\equiv \tanh\left ( \frac{\Delta\eta_{\ell\ell}}{2}
   \right)\; ,
\label{eq:barr}
\end{equation}
where $\Delta\eta_{\ell\ell}$ is the rapidity difference between the same 
  charge leptons. Notice that this quantity is invariant under longitudinal
boosts. We plot in Fig.\ \ref{fig:barr} the expected $\cos\theta^*_{\ell\ell}$
distributions for the SM background and the production of scalar and vector
resonances with mass 500 (700) GeV in the left (right) panel after cuts
(\ref{cuts1})--(\ref{cuts3}).  In obtaining this figure, we imposed that the
cross section for the production of spin-0 resonances is the same of the one
for spin-1 states. We also display the SM background alone to show its impact
on the distributions.

These figures clearly show that the $\cos \theta^*_{\ell\ell}$ distribution for
spin-1 and and spin-0 resonances are quite different and they can be used to
quantify the required integrated luminosity needed to discriminate between
them at a given CL. A simple $\chi^2$ analysis of the distributions shown in
Fig.\ \ref{fig:barr} yields a 99\% CL discrimination between spin-0 and spin-1
resonances of mass 500 (700) GeV for an integrated luminosity of 170 (215)
fb$^{-1}$, considering only the statistical errors.

In order to eliminate possible normalization systematics in the angular
distributions, we have also estimated the integrated luminosity needed to
decipher the spin of the new charged state by constructing an angular
asymmetry
\begin{equation}
  A_{\ell\ell} = 
\frac{\sigma(|\cos\theta^*_{\ell\ell}| < 0.5) - \sigma(|\cos\theta^*_{\ell\ell}| 
> 0.5) }
{\sigma(|\cos\theta^*_{\ell\ell}| < 0.5) + \sigma(|\cos\theta^*_{\ell\ell}| 
> 0.5) } \; .
\label{asym:emu}
\end{equation}
Considering only the statistical errors, this asymmetry allows a
$99$\% CL distinction between spin-0 and spin-1 resonances of mass 500
(700) GeV for an integrated luminosity of 440 (560) 
fb$^{-1}$.  With these choices of integrated luminosities, we have
$A_{\ell\ell}(\hbox{scalar}) = +0.104 \pm 0.05$ and
$A_{\ell\ell}(\hbox{vector}) = -0.07 \pm 0.05$, for $M_{V_1^\pm}=500$
GeV, and $A_{\ell\ell}(\hbox{scalar}) = -0.036\pm 0.06$ and
$A_{\ell\ell}(\hbox{vector}) = -0.27 \pm 0.06$, for $M_{V_1^\pm}=700$
GeV, where we have quoted only the statistical errors.

\begin{figure*}
       \centerline{\psfig{figure=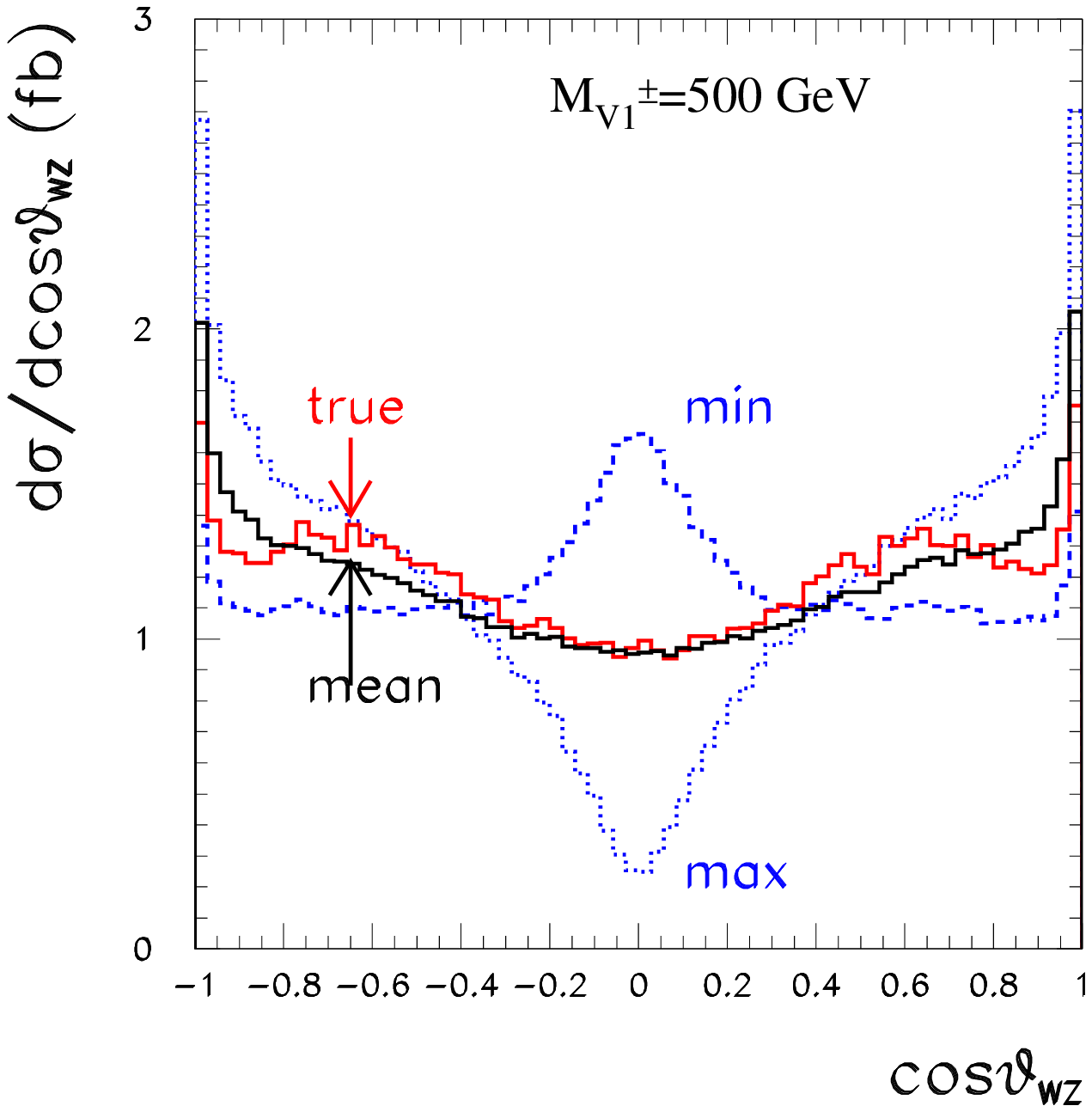,
width=0.55\textwidth,height=0.3\textheight}\hspace*{-1.5cm}
                   \psfig{figure=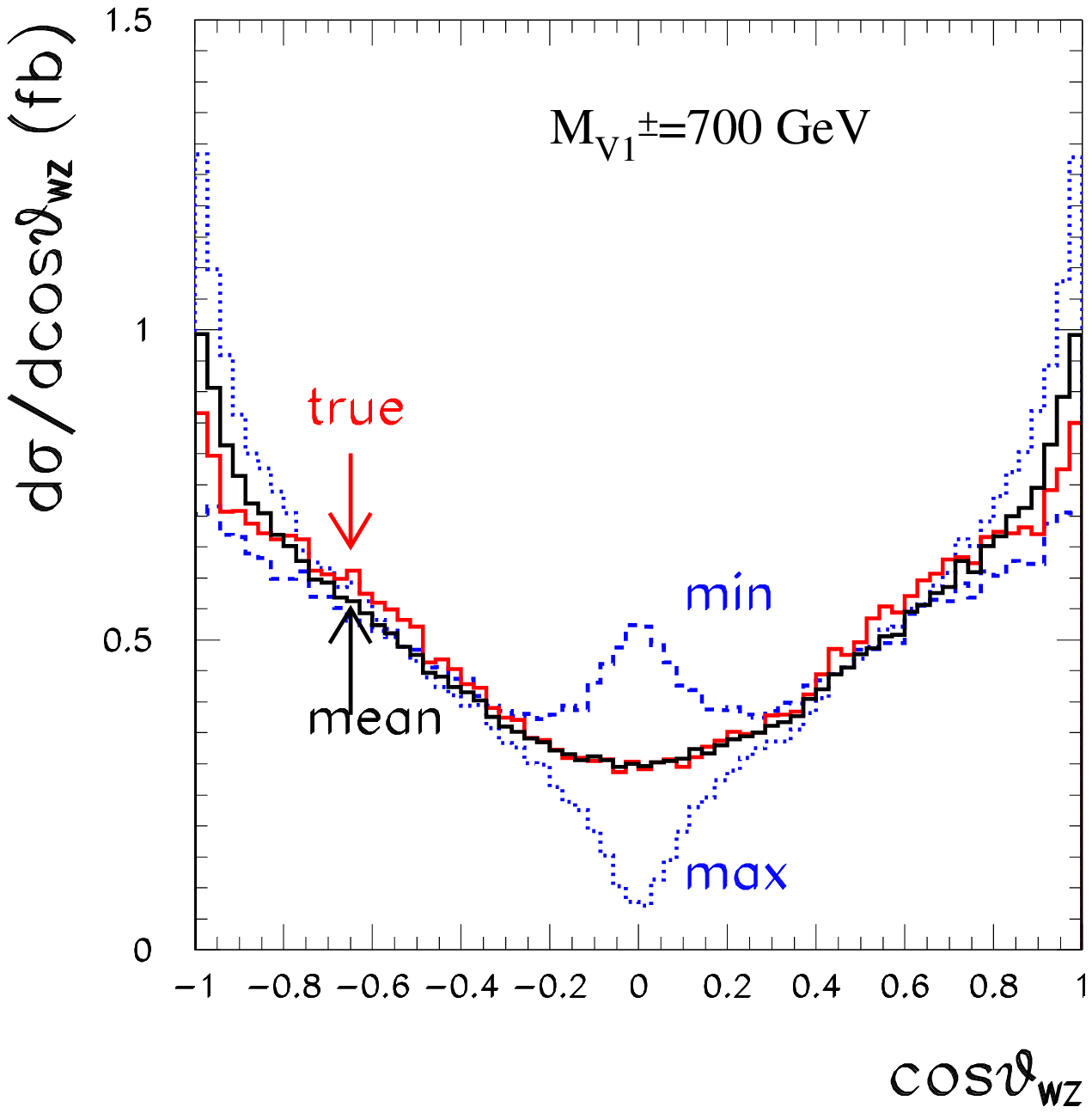,
width=0.55\textwidth,height=0.3\textheight}}
\caption{$\cos\theta_{WZ}$ distributions for a vector resonance of 500
  GeV (left panel) and 700 GeV (right panel). The dashed (dotted) line
  stands for the reconstructed distribution using the neutrino
  momentum that leads to the maximum (minimum) $WZ$ invariant
  mass. The solid lines represent the true distribution (upper line)
  and the one obtained averaging the reconstructed distributions with
  maximum and minimum $WZ$ invariant mass (lower line). }
\label{fig:wz}
\end{figure*}

We also studied the resolving power of the reconstructed $Z$ polar angle
($\theta_{WZ}$)
distribution evaluated in the $WZ$ center--of--mass frame. We display in
Fig.\ \ref{fig:wz} the $\cos\theta_{WZ}$ distribution for spin-1 charged
states after cuts (\ref{cuts1})--(\ref{cuts3}).  Since the reconstructed
neutrino momentum has a twofold ambiguity, there is also a twofold ambiguity in
the reconstructed $Z$ polar angle in the $WZ$ center--of--mass frame which
lead to the two distributions shown in the figure.  The dashed (dotted)
lines correspond to the reconstructed $Z$ polar angle distribution using the
neutrino momentum that leads to the maximum (minimum) $WZ$ invariant mass.  As
we can see, the two distributions differ appreciably for $\cos\theta_{WZ}$
close to zero. However, as shown in the figure, the average of the two
distributions has a better behavior in the central region of the detector
and resembles the true distribution.
Consequently, we have considered the average of the two reconstructed
distributions as discriminating observable. 

\begin{figure*}
       \centerline{\psfig{figure=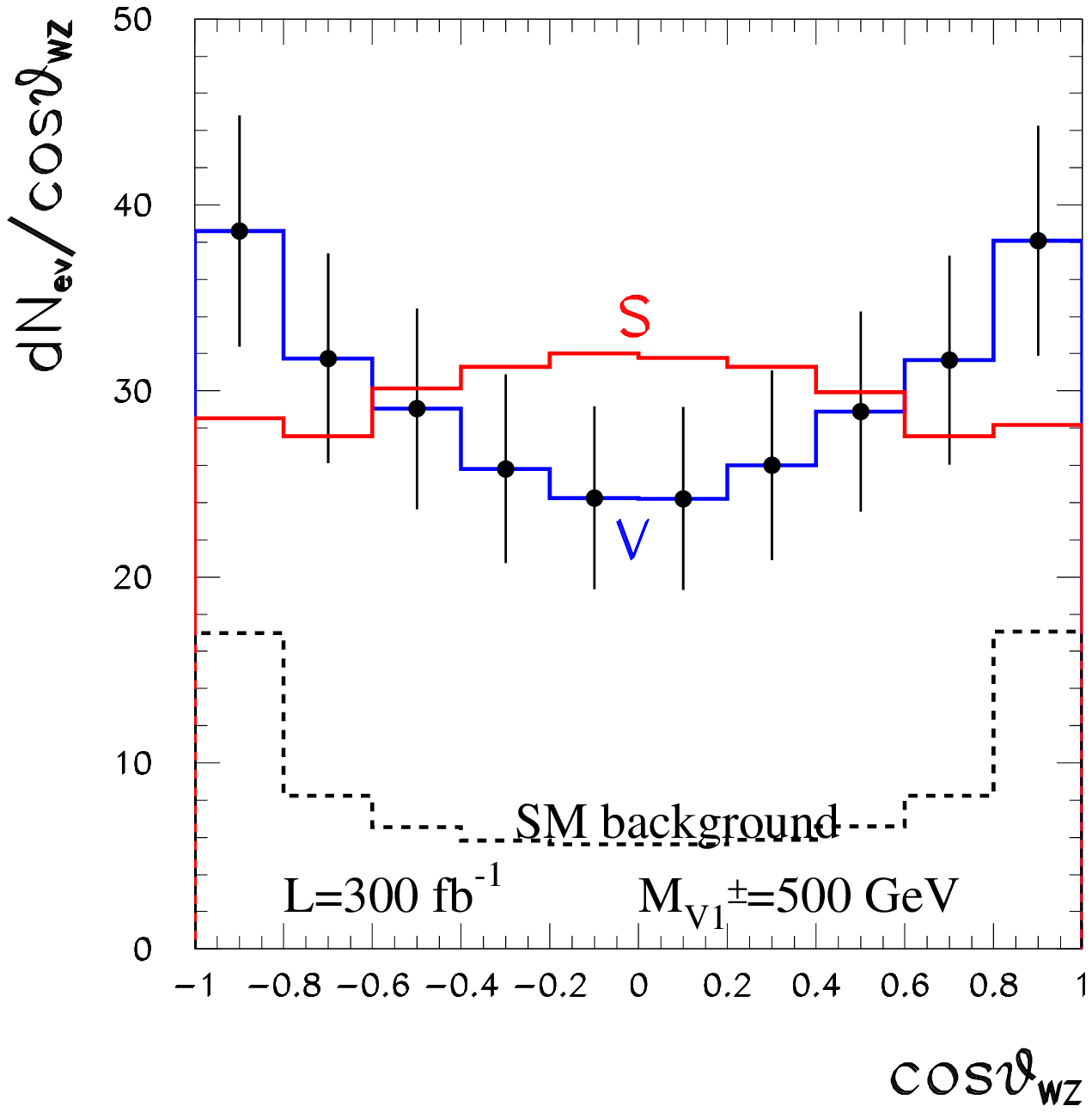,
width=0.55\textwidth,height=0.3\textheight}\hspace*{-1.5cm}
                   \psfig{figure=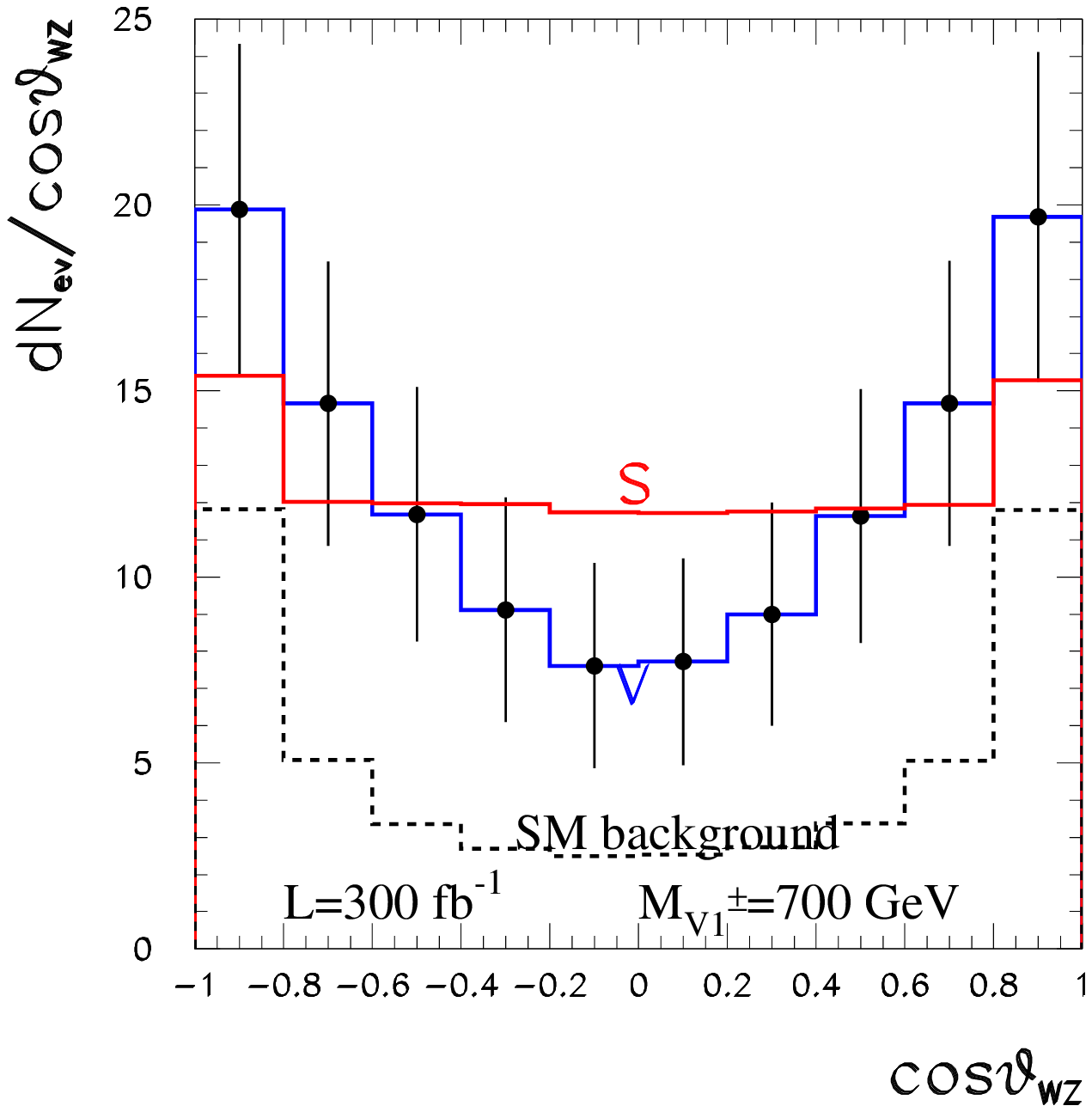,
width=0.55\textwidth,height=0.3\textheight}}
\caption{Same as Fig.\ \ref{fig:barr}, but for the $\cos\theta_{WZ}$
  distribution.}
\label{fig:coswz}
\end{figure*}

Fig.\ \ref{fig:coswz} depicts such averaged distributions for charged vector
and scalar resonances, where we are included the SM background prediction for
assessment of its impact on the spin determination.  Clearly, the production of
$V_1^\pm$ leads to more $WZ$ pairs produced at small polar angles while the
scalar resonance leads to more central events, as expected.  As above, in
order to quantify the discriminating power between the scalar and vector
productions we constructed the asymmetry
\begin{equation}
  A_{WZ} = 
\frac{\sigma(|\cos\theta_{WZ}| < 0.5) - \sigma(|\cos\theta_{WZ}| > 0.5) }
{\sigma(|\cos\theta_{WZ}| < 0.5) + \sigma(|\cos\theta_{WZ}| > 0.5) } \; .
\label{asym:wz}
\end{equation}
We find that for the new state mass of 500 (700) GeV, it is necessary
400 (550) fb$^{-1}$ to separate the two possibilities at 99\%
  CL.  With these choices of integrated luminosities, we have
$A_{WZ}(\hbox{scalar}) = +0.057 \pm 0.05 $ and $A_{WZ}(\hbox{vector}) =
-0.125 \pm 0.05$, for $M_{V_1^\pm}=500$ GeV, and $A_{WZ}(\hbox{scalar})
= -0.04 \pm 0.06$ and $A_{WZ}(\hbox{vector}) = -0.28 \pm 0.06$, for
$M_{V_1^\pm}=700$ GeV, where we have again quoted only the statistical
errors. 
Furthermore the use of a $\chi^2$
analysis of the $\cos\theta_{WZ}$ distribution is able to reveal the
spin of the new state at 99\% CL for an integrated luminosity of 
150 (220) fb$^{-1}$, for $M_{V_1^\pm} = 500$ (700) GeV.


\section{Conclusions}
\label{conclusions}

The observation of new charged vector resonances in Higgsless models decaying
into $WZ$ pairs can be carried out via weak boson production at the LHC
and their subsequent decays into charged
leptons~\cite{Birkedal:2004au,He:2007ge}.  Here we show how the LHC will be
able to determine the spin of these new states using two different
methodologies.  In the first method, only the observed charged leptons are used
to discriminate between spin-0 and spin-1 resonances using the variable
defined in Eq.~(\ref{eq:barr}). In this case, an integrated luminosity of 170
(215) fb$^{-1}$ is needed to establish the spin of the 500 (700) GeV resonance
at 99\% CL via a $\chi^2$ analysis of the $\cos\theta^*_{\ell\ell}$
distribution. On the other hand, the use of the asymmetry given by 
Eq.\ (\ref{asym:emu}) requires 440 (560) fb$^{-1}$ to determine the new 
resonance spin for a mass 500 (700) GeV. The second method is based on 
the two--folded reconstruction of the escaping neutrino momentum to obtain 
the $WZ$ polar angle distribution in its center--of--mass frame. This 
procedure requires a good understanding and calibration of the hadronic 
calorimeters, therefore, being subject to larger systematic
uncertainties.  We determined that the later method can distinguish
between spin-1 and spin-0 states at 99\% CL for integrated luminosities of 150
(220) fb$^{-1}$ for $M_{V_1^\pm}=500$ (700) GeV, respectively,  
when we perform a
$\chi^2$ fit of the $\cos\theta_{WZ}$ distribution.  If we use the asymmetry
defined in Eq.\ (\ref{asym:wz}) to perform the analysis, the integrated 
luminosities are 400 and 550 fb$^{-1}$, respectively. All results above
only account for statistical errors and  the 
inclusion of systematic uncertainties may render the efficiencies of the 
two methods rather similar.


\section*{Acknowledgments}

We thank G.\ Burdman for a careful reading of the manuscript.  This
work was supported in part by Conselho Nacional de Desenvolvimento
Cient\'{\i}fico e Tecnol\'ogico (CNPq) and by Funda\c{c}\~ao de Amparo
\`a Pesquisa do Estado de S\~ao Paulo (FAPESP); M.C.G-G is supported
by National Science Foundation grant PHY-0354776 and by Spanish Grants
FPA-2007-66665-C02-01, and FPA2006-28443-E.

\bibliographystyle{h-physrev}

\end{document}